# Directed Ligand Passage Over the Surface of Diffusion-Controlled Enzymes: A Cellular Automata Model

Mehrdad Ghaemi[1,2], Nasrollah Rezaei-Ghaleh[3], and Mohammad-Nabi Sarbolouki[3]

[1] Atomic Energy Organization of Iran, Deputy in Nuclear Fuel Production, Tehran, Iran
[2] Department of Chemistry, Tarbiat Moalem University, Tehran, Iran
{ghaemi@saba.tmu.ac.ir}
[3] Institue of Biochemistry and Biophysics, Tehran University, P.O. Box: 13145-1384, Tehran, Iran

**Abstract.** The rate-limiting step of some enzymatic reactions is a physical step, i.e. diffusion. The efficiency of such reactions can be improved through an increase in the arrival rate of the substrate molecules, e.g. by a directed passage of substrate (ligand) to active site after its random encounter with the enzyme surface. Herein, we introduce a cellular automata model simulating the ligand passage over the protein surface to its destined active site. The system is simulated using the lattice gas automata with probabilistic transition rules. Different distributions of amino acids over the protein surface are examined. For each distribution, the hydration pattern is achieved and the mean number of iteration steps needed for the ligand to arrive at the active site calculated. Comparison of results indicates that the rate at which ligand arrives at the active site is clearly affected by the distribution of amino acids outside the active side. Such a process can facilitate the ligand diffusion towards the active site thereby enhancing the efficiency of the enzyme action.

## 1 Introduction

The chemical machinery of many enzymes (proteins which act as catalysts of biochemical reactions) is so optimized that whenever a substrate molecule arrives at the active site of the enzyme and the enzyme-substrate complex is formed, the complex goes on the way to form the product instead of dissociating back to enzyme and substrate molecules. Under such conditions, the enzymes function at rates that depend on the diffusion-limited association of their substrates with the active site [1]. According to the Smulochowski equation, the maximum rate constant for the enzyme-substrate encounter in a solution is $k_{collision} = 4\pi Da$, where $D$ is the relative translational diffusion coefficient of enzyme and ligand molecules and $a$ is the sum of their effective radii [2]. For typical values of $D$ and $a$, $k_{collision}$ is of the order of $10^9$-$10^{10}$ $M^{-1}S^{-1}$ [3]. To account for any change in the diffusional encounter due to electrostatic interaction between the enzyme and substrate molecules, the Smulochowski equation is modified by introducing a dimensionless factor, $f$, [4] which does not however exceed 10 even for very favorable interactions [5]. Furthermore, regarding that only a small fraction of the enzyme surface is involved in the reaction, the Smulochowski equation should be modified further [6], then for typical enzymes, $k_{association}$ is not expected to exceed the value of $10^4$-$10^5$ $M^{-1}S^{-1}$ [7]. However, the association rate constants frequently exceed this theoretical limit [3]. Many models have been suggested to explain this discrepancy [8-10]. Herein, this problem is dealt with using a cellular automata model of ligand passage on the surface of enzyme molecules. It will be shown that the motion of ligand molecule on the surface of enzyme molecule is affected by different distributions of amino acid residues lying outside the active side of the enzyme; thereby their distribution may greatly facilitate arrival of the substrate molecule at the active site. This means that the effective area for the reaction may be larger than those considered merely based on geometrical assessment of the active site.

## 2 Method

In order to provide a simple representation of enzyme surfaces, a 27*27 grid of square cells with periodic boundary conditions was created (This is at the same order of magnitude as the number of amino acid residues occurring on the enzyme surfaces). The hydropathic profile of enzyme surfaces was represented assigning each cell an integer number between 0 and 8, corresponding to the hydropathic indices of amino acid residues occurring in those cells. The hydropathic indices were adopted from Kyte and Doolittle [11], then ranked from the most hydrophobic residues to the most hydrophilic ones in 9 scales, 0-8, i.e. 0 corresponds to the most hydrophobic residue- isoleucine- and 8 corresponds to the most hydrophilic one- arginine. Configuration of the entire system was then defined by the state values of all cells in the grid. Four configurations were examined in this study (Fig.1): A- a uniform configuration with all cells adopting a value of 4, i.e. occupied with a residue neither hydrophobic nor hydrophilic. The mean hydropathic index of this configuration is clearly 4 and its variance is zero. B- a random configuration with cells adopting numbers generated by a non- biased random method. This configuration has the same mean hydropathic index as the A configuration but displays a non-zero variance around the same mean. C- a configuration like A except that cells in diagonal, vertical and horizontal lines crossing the center of the grid adopt a value of 8, i.e. occupied with the most hydrophilic residues.

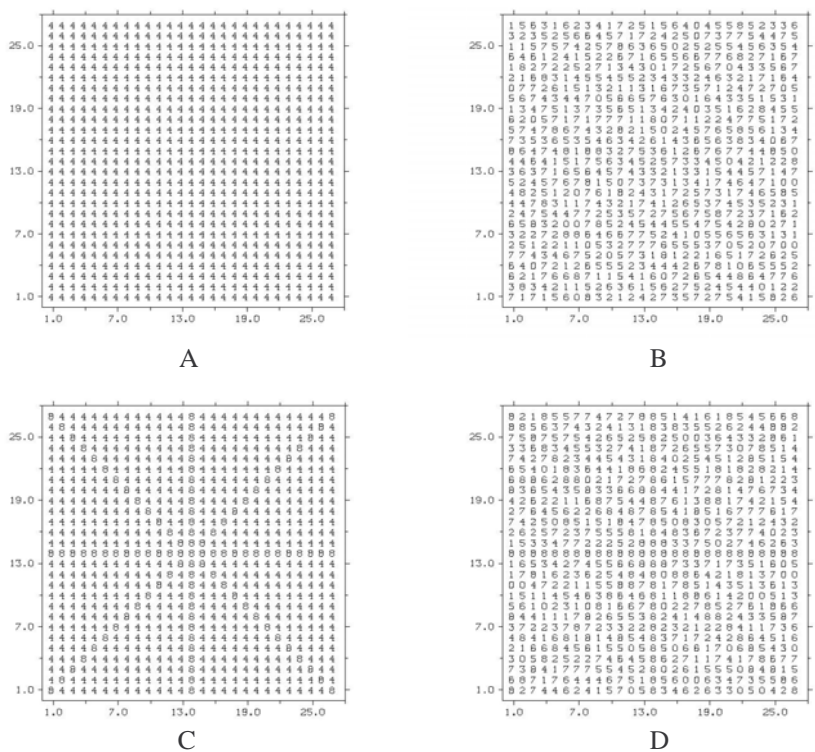

**Fig. 1.** Examined configurations of the entire system; See the text for a precise definition of each configuration.

D- a configuration like B except that cells in diagonal, vertical and horizontal lines crossing the center of the grid adopt a value of 8. The C and D configurations have the same mean hydropathic indices but show different non-zero variances around the same mean.

For each configuration mentioned above, water molecules were allowed to move around on the surface of grid and create a specific hydration pattern characteristic to that configuration. At first, each cell in the grid was assigned a random integer number between 0 and 99 indicating the number of water molecules, which exist in its vicinity. Then, the grid was partitioned into 81 3*3 blocks. At each step, the numbers of water molecules in each block were first added and then redistributed randomly among its 9 cells according to their hydropathic indices, i.e. the probability with which a water molecule occurs in a cell is directly proportional to its hydropathic index. Between steps, the blocks were displaced one cell toward the lower right direction. Iterations were continued until an almost constant hydration pattern was produced. It was judged qualitatively according to graphical views of the system and achieved typically after 50 iterations.

To assess the effect of each configuration with its characteristic hydration pattern on the rate at which substrate molecule arrives in the active site, we placed the substrate (ligand) molecule at the corner and the active site at the center of the grid and simulated the ligand motion over the surface of the grid. The ligand molecule could either be stationary or move along one of the four directions (up, down, right, left) with the same velocity (1 cell/iteration). Initially, the ligand molecule was assumed to be stationary. Then, it was allowed to move around on the surface of the grid according to the scheme presented by Boon et al. [12]. At each iteration, the ligand molecule first would take propagation according to its directed velocity. Then, during the subsequent redistribution step, the ligand molecule arrived in a new cell would lose its previous velocity thereby making its next decision merely on the basis of the local hydration pattern in its neighborhood: The probability with which the ligand molecule would adopt a specific directed velocity hence displace into a specific cell was directly proportional to the number of water molecules occurring in that cell. The system was iterated 5000 times for each configuration -at each iteration, different sets of random numbers were generated- and the average number of iterations needed for the ligand molecule to reach at the active site was calculated for them.

## 3 Results and Discussion

The hydration patterns, achieved for each of four configurations, are illustrated in Fig. 2. As expected, the distribution of water molecules over the grid obeys the hydropathic profile of each configuration. It is uniform in the case of configuration A and random in configuration B. In C and D configurations, the cells occurring on the diagonal, vertical and horizontal lines display more water molecules.

The average numbers of iterations required for the substrate to arrive at the active site were 2142, 2424, 1778 and 1984 for A, B, C and D configurations, respectively. It was significantly lower for configurations with specified diagonal, vertical and horizontal paths (C and D) than their counterpart configurations (A and B, respectively). A less prominent difference was observed between configurations with the same mean but different variances of hydropathic indices. The configurations with higher variances (B and D) showed higher average numbers of iterations required for the substrate to arrive at the active site than those with lower variances (A and C, respectively). These results support the idea that the hydropathic profile of enzyme surfaces away from the active site modifies the rate at which ligand arrives at the active site. Such an effect can be prominent for a chemically perfect enzyme, where the enzyme activity is diffusion-controlled.

The subject of ligand passage over the surface of proteins towards the active site has been extensively addressed in the literature. According to the current view, the ligand molecule temporarily resides in the vicinity of the protein surface after its encounter with the surface. Then, the ligand undergoes a two-dimensional walk toward the active site. It is generally believed that a two-dimensional surface diffusion to the active site can enhance the rate of a diffusion-controlled reaction [13]. Several models of two-dimensional surface diffusion have been presented. The directed passage of ligand toward the active site, as suggested by some of these models, enhances the rate of diffusion-controlled reactions even further. Some models suggest that the electrostatic field experienced by the ligand on the surface of protein may guide it toward the active site [14-15]. Some other models present the van der Waals interactions as the prominent force directing the ligand over the protein surfaces [10]. However, considering the effect of protein side chains on their vicinal water structure and the resultant changes on the local viscosity and diffusion coefficient, the ligand passage over the protein surface can be regarded as an anisotropic two-dimensional diffusion process. This anisotropy may provide preferred pathways for the ligand to arrive at the active site more rapidly. Using this approach, Kier et al. [16] have examined the ligand passage over hydrodynamic landscape of the protein by simulation via a cellular automata model. We hereby used a different cellular automata model to examine whether the hydropathic profile of protein surface can affect the rate at which the ligand arrives at the active site. The results presented here clearly indicate that the hydropathic profile of proteins can facilitate the ligand diffusion under some conditions. This is in accord with results obtained by Kier et al [16]. It can be interpreted that the existence of the preferred paths for ligand diffusion nearly converts a two-dimensional walk to a rather more efficient one-dimensional one. Indeed, in C and D configurations, the ligand molecule has spent most of its time within the preferred pathways moving toward or away from the active site. In comparison with the model presented by Kier et al. [16], our model may be judged to be simpler and more realistic. In addition, our model, due to its flexibility, may be more simply improved through the implementation of e.g. surface charge effects or detailed surface geometry. Such improvements are our current concerns regarding the model.

The activity of diffusion-controlled enzymes, which are chemically perfect, is limited by the physical step of ligand diffusion to the active site. It may be that such enzymes, through their evolution, have searched for ways to facilitate the ligand diffusion toward their active sites. The hydropathic profiles of the surfaces of such enzymes, when compared with other enzymes and/or among their variants, may provide some evidences regarding the existence of preferred profiles for ligand diffusion. Modification of these profiles lead to subtle changes in the activity of diffusion-controlled enzymes. Hence, the view presented in this study may provide some more reasonable tools for protein design and site-directed mutagenesis.

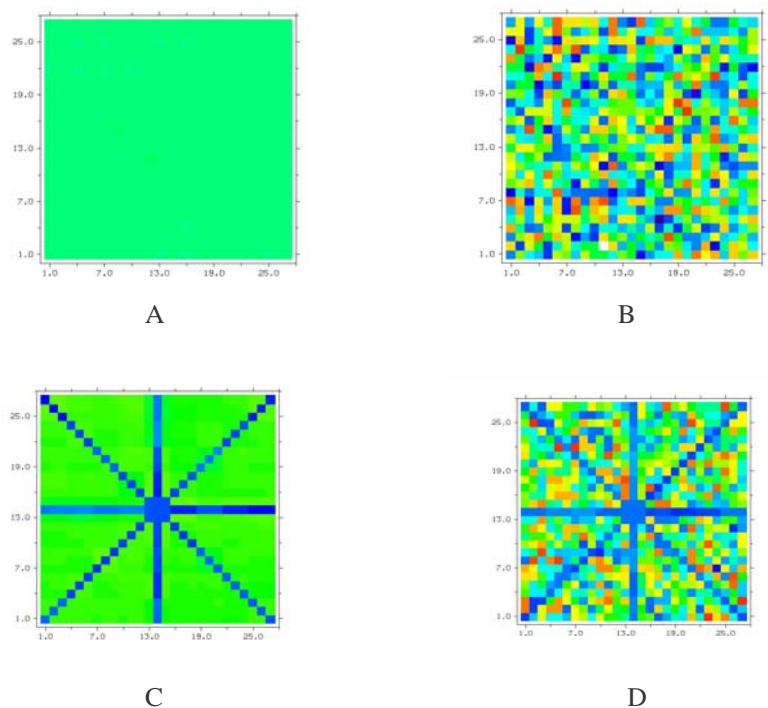

**Fig. 2.** Hydration patterns achieved for each examined configuration (A, B, C and D). The number of water molecules in each cell varies between 0 (dark red) and 99 molecules (dark blue). The intermediate numbers have been shown with colors of the spectrum between red and blue.

## 4 Conclusion

The presented cellular automata model of directed ligand passage over the surface of enzymes clearly shows that the hydropathic profile of enzyme surfaces can increase the rate at which the ligand molecule arrives in the enzyme active site. Such enhancing effect may be of functional importance in the case of enzymes with diffusion-controlled activity.

# References

1- Fersht, A.: Enzyme structure and mechanism. Freeman, New York (1985)

2- DeLisi, C.: The biophysics of ligand-receptor interactions. Q. Rev. biophys. 13 (1980) 201-230

3- Camacho, C.J., Weng, Z., Vajda, S., DeLisi, C.: Free energy landscapes of encounter complexes in protein-protein association. Biophys. J. 76 (1999) 1166-1178

4- von Hippel, P.H., Berg, O.G.: Facilitated target location in biological systems. J. Biol. Chem. 264 (1989) 675-678

5- Noyes, R.M.: Effects of diffusion rates on chemical kinetics. Prog. React. Kinet. 1 (1961) 129-160

6- Janin, J.: The kinetics of protein-protein recognition. Proteins: Struct. Funct. Genet. 28 (1997) 153-161

7- Schreiber, G., Fersht, A.R.: Rapid, electrostatically assisted association of proteins. Nature Struct. Biol. 3 (1996) 427-431

8- Berg, H.C., Purcell, E.M.: Physics of chemoreception, Biophys. J. 20 (1977) 193-215

9- Hasinoff, B.B.: Kinetics of acetylcholine binding to electric eel acetylcholine esterase in glycerol/water solvents of increased viscosity. Biochim. Biophys. Acta. 704 (1982) 52-58

10- Chou, K.C., Zhou, G.P.: Role of the protein outside active site on the diffusion-controlled reaction of enzyme. J. Am. Chem. Soc. 104 (1982) 1409-1413

11- Kyte, J., Doolittle, R.F.: A simple method for displaying the hydrophobic character of a protein. J. Mol. Biol. 157 (1982) 105-132

12- Boon, J.P., Dab, D., Kapral, R., Lawniczak, A.: Lattice gas automata for reactive systems. Physics Reports 273 (1996) 55-147

13- Welch, G.R.: The enzymatic basis of information processing in the living cell. Biosystems 38 (1996) 147-153

14- Radic, Z., Kirchhoff P.D., Quinin, D.M., McCommon, J.A., Taylor, P.: Electrostatic influence on kinetics of ligand binding to acetylcholinesterase. J. Biol. Chem. 272 (1997) 23265-23277

15- Wade, R.C., Gabdoulline, R.R., Ludeman, S.K., Lounnas, V.: Electrostatic steering and ionic tethering in enzyme-ligand binding: insights from simulations. Proc. Natl. Acad. Sci. U.S.A. 95 (1998) 5942-5949

16- Kier, L.B., Cheng, C.K., Testa, B.: A cellular automata model of ligand passage over a protein hydrodynamic landscape. J. Theor. Biol. 215 (2002) 415-426